\documentclass[prd,aps,showpacs,groupedaddress,eqsecnum,notitlepage,nofootinbib,twocolumn]{revtex4-2}
\usepackage{graphicx}
\usepackage[caption=false]{subfig}
\newcommand{\subfigimg}[3][,]{%
  \setbox1=\hbox{\includegraphics[#1]{#3}}%
  \leavevmode\rlap{\usebox1}%
  \rlap{\hspace*{-17pt}\raisebox{.5\baselineskip}{\small{#2}}}%
  \phantom{\usebox1}%
}
\usepackage{tikz,siunitx,mwe}
\usepackage{tensor}
\usepackage{epstopdf}
\usepackage{amsmath}
\usepackage{amsfonts}
\usepackage{slashed}
\usepackage{amssymb}
\usepackage{enumerate, color}
\usepackage{lipsum}
\usepackage{color}
\usepackage{graphicx,bm}
\usepackage[utf8]{inputenc}
\usepackage{eurosym}
\usepackage{scalerel}
\usepackage{float}
\usepackage{accents}
\usepackage[colorlinks=true,pdfstartview=FitV,linkcolor=blue,citecolor=blue,urlcolor=blue,breaklinks=true]{hyperref}
\usepackage{mathtools}
\newcommand{\be}{\begin{equation}}
\newcommand{\ee}{\end{equation}}
\newcommand{\ben}{\begin{eqnarray}}
\newcommand{\een}{\end{eqnarray}}
\newcommand{\bes}{\begin{subequations}}
\newcommand{\ees}{\end{subequations}}

\begin{document}
\title{Scalar fields and Lifshitz black holes from Derrick's theorem evasion}
\author{D. C. Moreira$^{1}$}
\author{F. A. Brito$^{1,2}$}
\author{J. C. Mota-Silva$^{3}$}
\affiliation{$^1$Unidade Acadêmica de Física, Universidade Federal da Campina Grande, 58109-970, Campina Grande, PB, Brazil}
\affiliation{$^2$Departamento de F\'isica, Universidade Federal da Para\'iba, 58051-970, Jo\~ao Pessoa, PB, Brazil}
\affiliation{$^3$Campus Salgueiro, Instituto Federal do Sertão Pernambucano, 56000-000, Salgueiro, PE, Brazil. }
\email{moreira.dancesar@gmail.com\\ fabrito@df.ufcg.edu.br\\
julio.mota@ifsertao-pe.edu.br}
\begin{abstract}
We study classical scalar fields in asymptotically Lifshitz spacetimes. By evading Derrick's theorem requiring the scalar potential to explicitly depend on the background coordinates, we induce a diffeomorphism invariance breaking and settle a formalism to find spatially localized solutions in the probe field limit. By inserting a backreaction equipped with a cosmological constant and a Maxwell term coupled to a dielectric function in the model, we find a Lifshitz black hole solution.
\end{abstract}
\maketitle
\section{Introduction}
Spatially localized structures modeled by classical scalars in field theory has brought several contributions to the understanding of topological solutions and their applications \cite{vilenkin2000cosmic,manton2004topological,vachaspati2006kinks}. In flat spacetime the existence of stable, nonzero finite-energy solutions in systems governed by standard covariant Lagrangians constituted only by scalar fields has hard restrictions arising from Derrick's theorem \cite{hobart1963instability,derrick1964comments}, which states that nontrivial solutions only exist in 1+1 dimensions in models where non-negative scalar potentials are equipped with some nonunitary set of degenerate minima. Spatially localized scalar field solutions are then trapped inside the topological sectors existing between consecutive degenerate minima and have a kinklike profile. 

There are some paths to evade Derrick's theorem in flat spacetimes and capture nontrivial scalar field solutions in arbitrary dimensions. One of these ways is presented in \cite{bazeia2003new}, where it's proposed to put aside general covariance by requiring the scalar potential to have an explicit dependence on the background coordinates. It generates nondynamical degrees of freedom in the action which implies that the energy-momentum tensor is no longer conserved in general. Studying models in this setup has provided several results involving topological solutions in different scenarios (recent related works are in \cite{bazeia2018dirac,bazeia2019configurational,casana2020bps,bazeia2021novel,bazeia2018magnetic,andrade2019first,bazeia2021configurational}). Generalization of Derrick's theorem for curved spacetimes has been proposed in some works \cite{radmore1978non,palmer1979derrick,carloni2019derrick}, pointing out that it is not possible to exist nontrivial probe scalar field solutions on flat or static, asymptotically flat spacetimes with more than 1+1 dimensions in relativistic scalar field models with standard covariant structure, but there are already paths for its evasion \cite{alestas2019evading,hartmann2020real,moreira2022analytical,moreira2022localized,morris2021radially,morris2022bps}, which leads to new tools on finding soliton-like structures in different scenarios. Some of these routes are also based on \cite{bazeia2003new}, but mainly on models with no backreaction since the diffeomorphism invariance breaking results in compatibility problems with the gravitational field equation when it comes to backreacting configurations. 

We are interested in working on radially symmetric backgrounds asymptotically approaching Lifshitz spacetimes \cite{kachru2008gravity} as
\begin{equation}\label{lifmetric}
\left.ds^2\right|_{r\to\infty}\simeq-\left(\frac{r}{\ell}\right)^{2z} dt^2+\left(\frac{\ell}{r}\right)^{2}dr^2+\left(\frac{r}{\ell}\right)^{2} dx^i dx^i,
\end{equation}
where $z$ denotes the {\it dynamical exponent}, read as a measure of the spacetime anisotropy since within its set of isometries the metric \eqref{lifmetric} is equipped with the nonrelativistic coordinate scaling \cite{taylor2016lifshitz}
\begin{equation}\label{lifscal}
\mathcal{D}_{z}: t\rightarrow \beta^z t, ~x^i\rightarrow \beta x^i~~\text{and} ~~ r\rightarrow r/\beta.
\end{equation}
Lifshitz geometries were introduced in \cite{kachru2008gravity} in the context of gauge/gravity duality to study strongly correlated nonrelativistic field theories presenting anisotropic scaling properties. For $z\neq1$ the boundary geometry \eqref{lifmetric} does not arise from scalar-tensor theories, but usually only emerges in the presence of massive vector fields \cite{taylor2008non} or in the context of Hor\"ava-Lifshitz gravity \cite{griffin2013lifshitz}. For $z=1$ we retrieve anti-de Sitter ($\text{AdS}_{D}$) spacetime. Additionally, black hole solutions asymptotically approaching Lifshitz spacetimes have emerged in models  where the temperature of the gravity side is used to involve the dual field theory in thermal bath \cite{taylor2008non,taylor2016lifshitz,hartnoll2009lectures,balasubramanian2009analytic} and since then several solutions of asymptotically Lifshitz black holes have been carried out (see, for instance \cite{ayon2010analytic,ayon2009lifshitz,bertoldi2009black,brito2020black,natsuume2018holographic,bazeia2015two,deveciouglu2011thermodynamics,gonzalez2011field,melnikov2019lifshitz,bravo2020thermodynamics,ayon2019microscopic}).

In this work we study scalar fields in asymptotically Lifshitz spacetimes both in probe limit and in the full backreacting setup by breaking general covariance (or diffeomorphism invariance). Initially, we are interested in capturing spatially localized structures in an effective model where the probe scalar field is placed on fixed, nonbackreacting geometries. In this regime, analytical and numerical solutions were found on the Lifshitz boundary geometry \eqref{lifmetric} \cite{moreira2022analytical} and on generic static, isotropic spacetimes, with applications in topological black holes and wormholes \cite{morris2021radially,morris2022bps,moreira2022localized}. In particular, in \cite{moreira2022analytical,moreira2022localized} a routine to capture nontrivial probe scalar solutions from first-order field equations on curved spacetimes has been introduced. Here we extend these studies to asymptotically Lifshitz spacetimes.

We insert backreaction in the model from a minimally coupled Einstein-Maxwell-scalar model equipped with a coordinate dependent dielectric function. In this scenario, in addition to losing coordinate invariance, we also have a compatibility problem in the field equations, since in the gravity sector the divergence of the Einstein tensor is zero (due to the contracted Bianchi identity which holds even in absence of general covariance) while the divergence of the energy momentum tensor in general is not due to contributions arising from the nondynamical degrees of freedom in the action. Consequently, we can only find solutions in the particular cases where the sum of contributions arising within the divergence of the energy momentum tensor due to the diffeomorphism invariance breaking cancel out, as extensively discussed, for example, in refs. \cite{bluhm2015spacetime,megevand2007scalar,anber2010breaking,bluhm2017gravity,bluhm2019gravity,bluhm2021gravity}. It leads us to a neutral topological Lifshitz black hole despite the presence of the Maxwell field, which is a net effect of the compatibility condition on the background dependent ingredients in the action. It is interesting to point out that although there is no prior correlation between the critical exponent and the nondynamic degrees of freedom, the latter vanish for $z\to1$, where one finds an AdS black hole. Thus, in this regime both relativistic scaling and general covariance are restored.

The work is organized as follows. In Sec. II we present the scalar field action with its respective field equations and develop the formalism to deal with probe scalars on Lifshitz spacetimes, where we study the formation of soliton-like structures. In Sec. III we insert backreaction in the model and find a Lifshitz black hole solution, whose thermodynamics is analyzed in Sec. IV. In Sec. V we make final remarks about the study we present.
\section{Scalar fields in the probe limit}
\subsection{General setup}
The action for the model we study in this section is given by
\begin{equation}\label{action}
S_{\left(\phi\right)}=\int d^{D} x\sqrt{-g}\left(-\frac{1}{2}\nabla_a \phi\nabla^a\phi-V(x,\phi)\right),
\end{equation}
where $g=\text{det}\left(g_{ab}\right)$ is the metric determinant, $\phi(x)$ represents a scalar field that self-interacts through the scalar potential $V(x,\phi)$, which explicitly depends on the background coordinates $x^a$, $a=0,1,\cdots,D-1$. The field equation and the energy-momentum tensor derived from the action \eqref{action} are 
\begin{subequations}
\begin{eqnarray}
\Box \phi&=&\frac{\partial V}{\partial\phi} \label{fieldeq},\\
\Theta_{ab}&=&\nabla_a\phi\nabla_b\phi-\frac{1}{2}g_{ab}\left(\nabla\phi\right)^2-g_{ab}V(x,\phi),\label{emt}~~
\end{eqnarray}
\end{subequations}
respectively, where $\Box=g^{ab}\nabla_a\nabla_b$ is the d'Alembertian operator. By requiring the action to explicitly depend on the background coordinates, we explicitly break the diffeomorphism invariance of the model, which implies - in particular - that the energy-momentum tensor is no longer conserved, since
\begin{equation}
    \nabla_a \Theta^a_{~b}=\partial_b V(x,\phi)\neq 0.~~\left(\text{in general}\right)
\end{equation}
The explicitly coordinate dependence can be read as the presence of nondynamical degrees of freedom in the action, which induces part of the self-interaction energy of the scalar field to be dissipated on the background geometry in form of energy and momentum, but for now not in quantity enough to change the background geometry.

We are interested in finding spatially localized structures in the probe limit where the scalar field is placed on fixed, static geometries describing asymptotically Lifshitz spacetimes generically written as
\begin{equation}\label{backmetric}
ds^2=-\left(\frac{r}{\ell}\right)^{2z}e^{2\nu}dt^2+\left(\frac{\ell}{r}\right)^{2}e^{2\lambda}dr^2+\left(\frac{r}{\ell}\right)^{2}\hat{\sigma}_{ij}dx^i dx^j,~~
\end{equation}
with $\nu=\nu(r)$, $\lambda=\lambda(r)$ and $(x^0,x^1)=(t,r)$. The standard Lifshitz spacetime \eqref{lifmetric} is recovered in cases where the limit $\left(e^{2\nu},e^{2\lambda}\right)_{r\to\infty}=(1,1)$ is satisfied along with the identification $\hat{\sigma}_{ij}=\delta_{ij}$, but here we assume a larger setup where the metric $\hat{\sigma}_{ij}(x^k)$ depends on the remaining coordinates $x^i,~i=2,\cdots,D-1,$ describing a transverse $(D-2)$-dimensional Einstein manifold $\hat{\Sigma}_{\gamma}$ with Ricci tensor $\hat{R}_{ij}=\gamma\hat{\sigma}_{ij}$ and $\gamma=0,\pm1$, hence $\hat{\Sigma}_{\gamma}$ can hold $S^{D-2}$ (for $\gamma=1$), $\mathbb{R}^{D-2}$ (for $\gamma=0$) or $H^{D-2}$ (for $\gamma=-1$) topologies. 

The background metric \eqref{backmetric} supports a time-like Killing vector $\xi=-\partial_t$, which can be used to build a conserved current related to the probe field as $J^{a}=\Theta^{ab}\xi_b$ and define the energy of the scalar as
\begin{equation}
\label{consch}
E(\xi)=-\int_{\Sigma }d^{D-1} x\sqrt{|h|} n_a\xi_b \Theta^{ab},
\end{equation}
where $\Sigma$ denotes a $(D-1)$-dimensional manifold defined at fixed time, equipped with unit normal vector $n_a=-\left(\frac{r}{\ell}\right)^{z}e^{\nu}\delta_{a0}$ and induced metric $h_{pq}$, with $h=\text{det}\left(h_{pq}\right)$ and $p,q=1,\cdots, D-1$. For simplicity, we assume that the scalar field and potential only depends on the radial coordinate, i.e., $\phi=\phi(r)$ and $V(x,\phi)=V(r,\phi)$. Opting for a static configuration of the model is important here because it allows us to ensure that there is no energy flux from the field to the background geometry. In this way, the energy of the field is conserved even in the presence of a gradient term of the scalar potential along the radial coordinate.

Under these considerations, the field equation \eqref{fieldeq} becomes
\begin{equation}\label{2ordereq}
\left(\frac{\ell}{r}\right)^{D+z-3}e^{-\left(\nu+\lambda\right)}\frac{d~}{dr}\left(\left(\frac{r}{\ell}\right)^{D+z-1}e^{\nu-\lambda}\frac{d\phi}{dr}\right)=\frac{\partial V}{\partial\phi}.
\end{equation}
The above equation is second order and in general hard to solve due to possible nonlinearities arising from the metric and scalar potential functions. For regular spatially localized solutions with finite energy the appropriate boundary conditions are
\begin{subequations}
\label{boundcond}
\begin{eqnarray}
   \phi(r\to r_0)&=&\phi_{0},~~\phi(r\to\infty)=\phi_{\infty},\\[4pt]
\lim_{r\to r_0}\left|\frac{d\phi}{dr}\right|&<&\infty,~~\lim_{r\to \infty}\frac{d\phi}{dr}=0,
\end{eqnarray}
\end{subequations}
where $\phi_0$ and $\phi_\infty$ are constants specified in each scalar field model and $r_0\geq 0$ denotes a possible lower bound on the radial coordinate range, depending on the background geometry one uses. 
\subsection{Minimal energy setups for probe scalars}
A possible route for finding spatially localized field solutions within the system described so far is by reducing the order of the field equation \eqref{2ordereq}, since it can relieve some difficulties when dealing with its nonlinearities. In order to capture suitable first-order equations, we look up for minimal energy field solutions. To begin with, we note that $-n_a\xi_b \Theta^{ab}=-\left(\frac{r}{\ell}\right)^{z}e^{\nu}\Theta^{0}_{~0}$ and that $\Theta^0_{~0}$ can be rearranged as
\begin{equation}\label{enrg}
 -\Theta^{0}_{~0}=\frac{1}{2}\left(e^{-\lambda}\frac{r}{\ell}\frac{d\phi}{dr}\mp\sqrt{2V}\right)^2\pm e^{-\lambda}\frac{r}{\ell}\frac{d\phi}{dr}\sqrt{2V}.
\end{equation}
By inserting these expressions in the energy of the scalar field in \eqref{consch} one can improve the bound
\begin{equation}
    E(\xi)\geq\pm\int_{\Sigma} d^{D-1} x\sqrt{|h|}e^{\nu-\lambda}\left(\frac{r}{\ell}\right)^{z+1}\frac{d\phi}{dr}\sqrt{2V},
\end{equation}
which is saturated for cases where the first-order differential equation
\begin{equation}\label{1oeq}
\frac{d\phi}{dr}=\pm e^\lambda\frac{\ell}{r}\sqrt{2V}
\end{equation}
is satisfied, leading us to minimal energy field solutions. A direct checking proves that scalar fields solving equation \eqref{1oeq} also satisfy the second-order equation \eqref{2ordereq}. A consequence of the first-order equation \eqref{1oeq} is that its solutions must come in pairs. Furthermore, since the sign of the derivative does not change, such solutions are monotonic and, in accordance with the boundary conditions \eqref{boundcond}, must present kinklike profiles. We use kink for the positive sign and antikink for the negative sign in Eq. \eqref{1oeq}.

The minimal energy is given by $E_{min}=\text{min}\left\{E(\xi)\right\}$ and by using relation \eqref{1oeq} one can express it as
\begin{equation}\label{bpsdef}
    E_{min}=\omega^{(\gamma)}_{D-2}\int_{r_0}^\infty dr e^{\nu-\lambda}\left(\frac{r}{\ell}\right)^{D+z-1}\left(\frac{d\phi}{dr}\right)^2,
\end{equation}
where $\omega^{(\gamma)}_{D-2}=\int d^{D-2}x\sqrt{|\hat{\sigma}|}$ is the worldvolume of the transverse space $\hat{\Sigma}_{\gamma}$, with $\hat{\sigma}=\text{det}\left(\hat{\sigma}_{ij}\right)$. Moreover, one can also use the equation \eqref{1oeq} to rewrite the energy density of the scalar field and shows that it automatically satisfies the weak energy condition
\begin{equation}\label{wec}
\rho=e^{-2\lambda}\left(\frac{r}{\ell}\right)^2\left(\frac{d\phi}{dr}\right)^2=\Theta_{ab}\xi^a\xi^b\geq0,
\end{equation}
which indicates that the scalar fields we are treating here are well behaved since its energy density measured by observers on time-like curves on the background geometry is always nonnegative \cite{kontou2020energy}. 

We can distinguish the kink solution from the antikink solution by associating the field values on the spacetime boundaries with conserved quantities. Here, we follow a routine similar to the one recently settled in \cite{moreira2022analytical,moreira2022localized} and build such a charge from the 1-form
\begin{equation}
B=B_a dx^a=e^{\nu+\lambda}\Delta\phi(r)\zeta(r)dt, 
\end{equation}
with $\Delta\phi(r)=\phi(r)-\phi_0$ and
\begin{eqnarray}\label{intpot}
   \zeta(r)=\left\{
\begin{array}{rcl}
&-\frac{r}{\alpha-1}\left(\frac{\ell}{r}\right)^{\alpha}&, ~~\text{if}~~ \alpha\neq 1,\\[3pt]
&\ell\ln r&, ~~\text{if}~~ \alpha=1,
\end{array}
\right.
\end{eqnarray}
where for simplicity we write $\alpha=D-z-1$.
The vector $B_a$ is used to define the anti-symmetric tensor $f_{ab}=\partial_aB_b-\partial_b B_a$, which is necessary to build the auxiliary conserved current $\widetilde{J}^{a}=\nabla_bf^{ab}$. Applying Gauss's Law on the current $\widetilde{J}^a$, one finds the conserved charge
\begin{subequations}
\begin{eqnarray}
\label{q1f}
Q_{\small{\Delta\phi}}&=&-\oint_{\partial\Sigma_\infty} d^{\small{D-2}}x \sqrt{|h^{(2)}|} n_a s_b f^{ab},\\[3pt]
&=&\omega^{(\gamma)}_{D-2}\Delta\phi_{\infty}\!\lim_{r\to\infty}\!\left(1+\left(\frac{r}{\ell}\right)^{\!\alpha}\!\left(\nu+\lambda\right)'\!\zeta(r)\right)\!,~~~~~~~~
\end{eqnarray}
\end{subequations}
where $\Delta\phi_{\infty}=\phi_{\infty}-\phi_0$ and  $s^a=e^{-\lambda}\frac{r}{\ell}\delta^{a}_{1}$ denotes the unit normal vector to the boundary surface $\partial\Sigma$ defined at fixed $r$, with induced metric $h^{(2)}_{ij}=\left(\frac{r}{\ell}\right)^2\hat{\sigma}_{ij}$ and taken at spatial infinity $\left(\partial\Sigma_{\infty}\right)$.

The last important ingredient to setup this formalism is given through the insertion of an auxiliary function $W(\phi)$ as follows:
\begin{equation}\label{1ordereq}
\frac{d\phi}{dr}=\pm  e^{-(\nu-\lambda)}\left(\frac{\ell}{r}\right)^{D+z-1}\frac{dW}{d\phi}.
\end{equation}
Under this choice the above first-order equation is - in principle - calculable by using standard calculus techniques, depending on the choice we made for the background geometry and the scalar potential, which now can be factored out by separating the field-dependent term from the coordinate-dependent one, leading to
\begin{equation}\label{potmodel}
V(r,\phi)=\frac{1}{2}e^{-2\nu}\left(\frac{\ell}{r}\right)^{2(D+z-2)}\left(\frac{dW}{d\phi}\right)^2,
\end{equation}
once we require compatibility between equations \eqref{1oeq} and \eqref{1ordereq}. In order to avoid possible divergences arising from the metric functions and ensure the compatibility of the scalar potential and the field equation \eqref{1ordereq} with the boundary conditions \eqref{boundcond} we also require that the scalar potential satisfy $V(r_0,\phi_0)=0=\lim_{r\to\infty}V(r,\phi)$, which implies that the boundary values  $(\phi_0,\phi_\infty)$ have to be extremes of $W(\phi)$. Another advantage of the presented formalism is that the minimal energy now depends only on the boundary values of the auxiliary function $W(\phi)$,
\begin{equation}\label{ebps}
    E_{min}=\omega^{(\gamma)}_{D-2}|\Delta W|,
\end{equation}
where $\Delta W=W(\phi_\infty)-W(\phi_0)$. In this way, we can find the energy of the scalar even if we do not know the exact solution of the field equation, since it's enough to know the auxiliary function related to the model and the scalar field boundary values.
\subsection{Example}
\begin{figure}[t!]	
		\centering
\includegraphics[width=8cm]{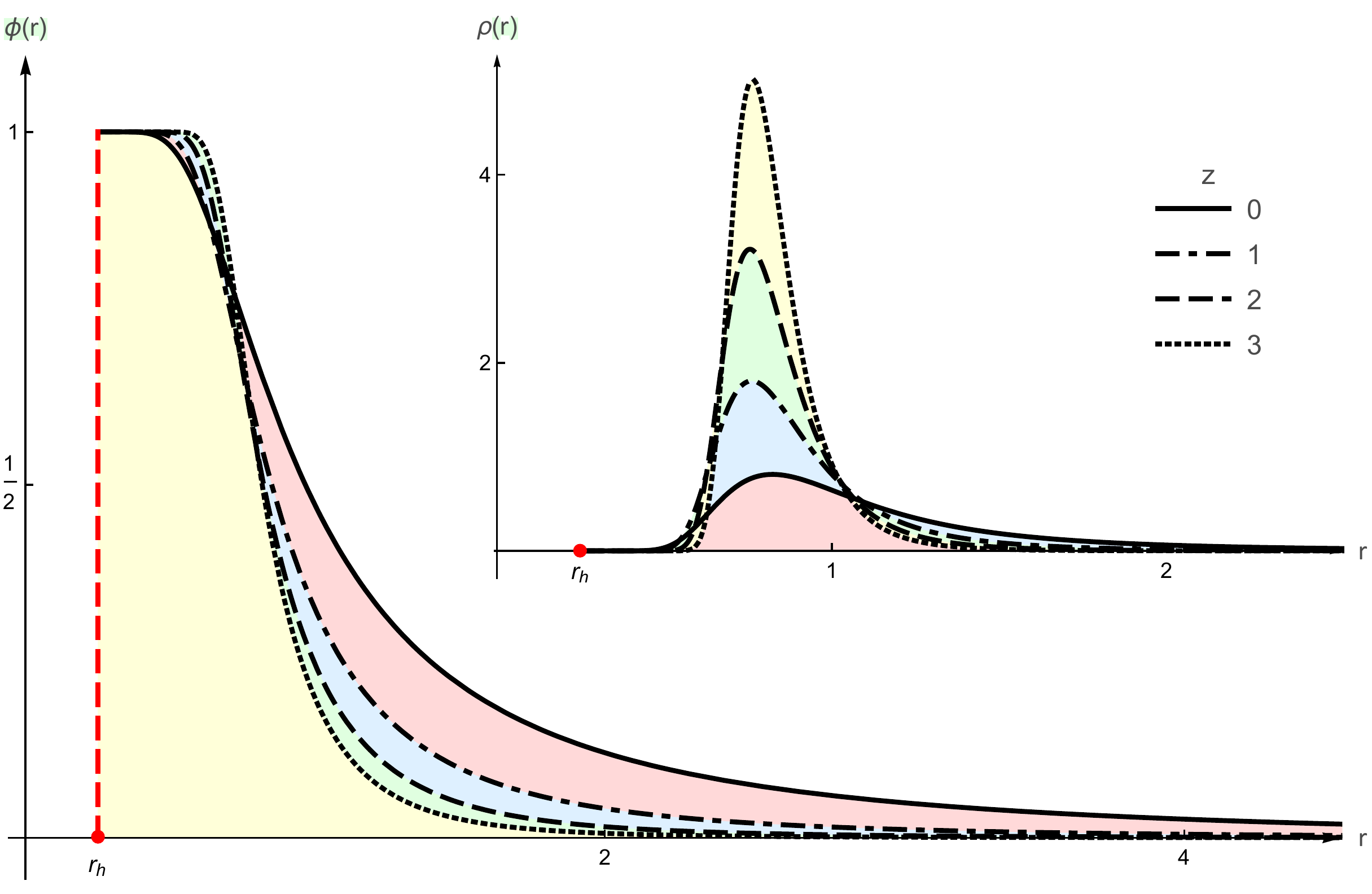}\\
\vspace{2mm}
\caption{Numerical solution  of the Eq. \eqref{fieldeqt} and its energy density \eqref{edens1} for $D=3+1$, $\ell=1$, $\gamma=0$, $r_h=1/4$ and some different values of $z$.}
\label{fig1}
\end{figure}
Now let's discuss an illustrative example where the ideas discussed so far can be applied providing analytical and numerical field solutions. We use as auxiliary function $W(\phi)=\phi-\phi^3/3$ and for background geometry we pick out the uncharged scenario of the Lifshitz topological black hole solution found in \cite{dayyani2018critical}, 
\begin{equation}
    e^{2\nu}=e^{-2\lambda}=1-\frac{2m}{r^{\beta-1}}+\gamma \widetilde{a}^2\left(\frac{\ell}{r}\right)^2,
\end{equation}
with $\widetilde{a}=(D-3)/(\beta-3)$, $\beta=D+z-1$ and $2m=r_h^{\beta-1}+\gamma \widetilde{a}^2\ell^2r_h^{\beta-3}$, where $r_0=r_h$ denotes its event horizon, which exists for any possible value of $\gamma$ in cases where $\beta-1>0$. With these ingredients the self-interaction potential \eqref{potmodel} becomes
\begin{equation}
    V(r,\phi)=\frac{1}{2}\left(\frac{\ell}{r}\right)^{2\left(\beta-1\right)}\frac{\left(1-\phi^2\right)^2}{1-\frac{2m}{r^{\beta-1}}+\gamma \widetilde{a}^2\left(\frac{\ell}{r}\right)^2},
\end{equation}
and the first-order field equation \eqref{1ordereq} is here given by
\begin{equation}\label{fieldeqt}
    \frac{d\phi}{dr}=\pm\left(\frac{\ell}{r}\right)^{\beta}\frac{1-\phi^2}{1-\frac{2m}{r^{\beta-1}}+\gamma \widetilde{a}^2\left(\frac{\ell}{r}\right)^2}.
\end{equation}
The above equation has no general solution for all possible values of $z$, but it leads to scalar fields asymptotically behaving as
\begin{equation}
    \phi(r\to\infty)\simeq\mp\frac{\ell}{\beta-1}\left(\frac{\ell}{r}\right)^{\beta-1}+\mathcal{O}\left(\frac{1}{r^{2\left(\beta-1\right)}}\right),
\end{equation}
where we consider $\phi_{\infty}=0$ in order to completely suppress the presence of the field in regions far from the black hole horizon, where the Lifshitz background  \eqref{lifmetric} is dominant. Analytical solutions emerge in cases where $z=3-D$. In these cases we have event horizon given by $r_h=m+\widetilde{m}$, where $\widetilde{m}^2=m^2-\gamma (D-3)^2\ell^2$ and the scalar field  - in terms of the parameters $m$ and $\widetilde{m}$ - becomes
\begin{eqnarray}\label{fieldsol}
   \phi(r)=\left\{
\begin{array}{rcl}
&\pm\tanh\left(\frac{\ell^2}{2\widetilde{m}}\ln\left(\frac{1-(m+\widetilde{m})/r}{1-(m-\widetilde{m})/r}\right)\right)&, ~\text{if}~ \widetilde{m}> 0,\\[8pt]
&\mp\tanh\left(\frac{\ell^2}{r-m}\right)&, ~\text{if}~ \widetilde{m}=0.
\end{array}
\right.
\end{eqnarray}
The energy density of the field is
\begin{equation}\label{edens1}
    \rho=\left(\frac{\ell}{r}\right)^{2\left(\beta-1\right)}\frac{\left(1-\phi^2\right)^2}{1-\frac{2m}{r^{\beta-1}}+\gamma \widetilde{a}^2\left(\frac{\ell}{r}\right)^2},
\end{equation}
and has its profile depicted in Fig.\eqref{fig1} along with a numerical field solution of \eqref{fieldeqt} for different values of $z$.  Note that it has a spatially localized shape where the increasing of the $z$-parameter accentuates the rate the field is asymptotically suppressed, in addition to making the field density more concentrated in the sectors near the horizon. As required, the field is regular throughout the region outside the event horizon and has a kinklike profile. In the vicinity of the horizon the field assumes one of the extremes of $W(\phi)$ and the energy density strongly approaches zero, delimiting the region where the curvature effects are dominant over any self-interacting response of the field. Finally, the energy in \eqref{ebps} and the conserved charge \eqref{consch} are evaluated as 
\begin{equation}
\left(E_{min},~Q_{\small{\Delta\phi}}\right)=\omega^{(\gamma)}_{D-2}\left(2/3,\pm1\right),
\end{equation}
where the $``+"$ ($``-"$) sign is used to characterize the kink (antikink) solution. 
\subsection{Radial stability}
The radial stability of the solutions found from the model we are discussing in this section is given by considering periodic fluctuations around the static solution as $\phi(r,t)=\phi(r)+e^{i\omega t}\psi(r)$ in the field equation \eqref{2ordereq}, which provides the stability equation
\begin{eqnarray}\label{steq}
\left(-\Box+\left.\frac{\partial^2 V}{\partial\phi^2}\right|_{\phi(r)}\right)\psi(r)=\omega^2\left(\frac{\ell}{r}\right)^{2z}e^{-2\nu}\psi(r).~~~~~
\end{eqnarray}
After some manipulations, one can rewrite the Eq. \eqref{steq} as a Sturm-Liouville problem,
\begin{equation}\label{stmlvl}
   \left(-\frac{d~}{dr}\left(p(r)\frac{d~}{dr}\right)+q(r)\right)\psi=\omega^2\left(\frac{r}{\ell}\right)^{\alpha-2}e^{-(\nu-\lambda)}\psi,
\end{equation}
where
\begin{equation}
p(r)=\left(\frac{r}{\ell}\right)^{\beta}e^{\nu-\lambda} ~~\text{and}~~
q(r)=\left(\frac{r}{\ell}\right)^{\beta-2}e^{\nu+\lambda}\left.\frac{\partial^2V}{\partial\phi^2}\right|_{\phi(r)},
\end{equation}
with, again, $\beta=D+z-1$. Its compatible inner product is
\begin{equation}\label{inner}
    \langle \psi_m,\psi_n\rangle=\int_{r_o}^{\infty}dr \left(\frac{r}{\ell}\right)^{\alpha-2}e^{-\left(\nu-\lambda\right)}\Bar{\psi}_m(r)\psi_n(r).
\end{equation}
Equation \eqref{stmlvl} can be factorized as $S^{\dagger}S\widetilde{\psi}=\omega^2\widetilde{\psi}$ where $\widetilde{\psi}=e^{\nu-\lambda}\psi$ and 
\begin{subequations}
\begin{eqnarray}
S^{\dagger}&=&\left(\frac{r}{\ell}\right)^{z+1}e^{\nu-\lambda}\left(\frac{d~}{dr}+\mathcal{W}(r)+\xi(r)\right),\\[3pt]
S~&=&\left(\frac{r}{\ell}\right)^{z+1}e^{\nu-\lambda}\left(-\frac{d~}{dr}+\mathcal{W}(r)\right),
\end{eqnarray}
\end{subequations}
with
\begin{subequations}
\begin{eqnarray}
\mathcal{W}(r)&=&\frac{z+1}{r}\!+\!\frac{W_{\phi\phi}}{r^{D-2}}\left(\frac{\ell}{r}\right)^{\!z+1}\!\!\!e^{-(\nu-\lambda)}+\!\left(\nu-\lambda\right)'\!,~~~~~~~~\\[3pt]
\xi(r)&=&2\left(\nu-\lambda\right)'-\frac{D-2 (z+2)}{r}.
\end{eqnarray}
\end{subequations}
One can show that the operators $S^\dagger$ and $S$ are adjoint under the inner product \eqref{inner} \cite{hounkonnou2004factorization}, which implies that the Sturm-Liouville problem \eqref{stmlvl} have no states with negative energy, which ensures the radial stability of the solutions found. In particular, the zero-mode (ground-state) is extracted from equation $S\widetilde{\psi}_0=0$, resulting in
 \begin{equation}
    \psi_0=c\exp{\left(\int dr \frac{W_{\phi\phi}}{r^{D-2}}\left(\frac{\ell}{r}\right)^{z+1}e^{-(\nu-\lambda)}\right)},
\end{equation}
where $c$ denotes a normalization constant.

\section{Lifshitz black hole solution}
Now we consider backreaction effects in the system. We use the action of an Einstein-Maxwell-scalar model given by
\begin{equation}\label{1action}
S=S_{\left(\phi\right)}+\frac{1}{2}\int d^{D}x\sqrt{-g}\left(R-2\Lambda-\varepsilon(x)F_{ab}F^{ab}\right),
\end{equation}
where $S_{\left(\phi\right)}$ denotes the scalar action \eqref{action}, $R$ represents the Ricci scalar, $F_{ab}=\nabla_a A_b-\nabla_b A_a$ is the Maxwell tensor and $\Lambda=-\left(D-2\right)\left(D+3z-4\right)/2\ell^2$ plays the role of a cosmological constant. The Maxwell field is coupled with a coordinate-dependent function $\varepsilon(x)$, denoting a nondynamical background field which induces another diffeomorphism breaking in the model and plays the role of an effective dielectric \cite{lee1991bogomol,bazeia1992vortices}. In usual scalar-tensor models presenting this type of coupling the effective dielectric depends only on the scalar field (a necessary condition to ensure general covariance in these scenarios) and it has recently been used in modelling scalarization mechanisms for black holes (see, for instance, \cite{herdeiro2018spontaneous,javed2019effect,guo2021scalarized,herdeiro2021spontaneous,astefanesei2020higher,zhang2022evolution,yao2021scalarized,konoplya2019analytical,fernandes2020einstein, fernandes2019spontaneous} and references therein). 

The field equations derived from the action \eqref{1action} are 
\begin{subequations}
\begin{eqnarray}
\label{ee1}\Box \phi-\frac{\partial V}{\partial\phi}&=&0,\\[3pt]
\label{ee2}\nabla_a \left(\varepsilon(x) F^{ab}\right)&=&0,\\[3pt]
\label{ee3}\mathcal{E}_{ab}=G_{ab}+\Lambda g_{ab}-T_{ab}&=&0,
\end{eqnarray}
\end{subequations}
where $G_{ab}$ denotes the Einstein tensor and 
\begin{equation}\label{emt1}
T_{ab}=\Theta_{ab}+\varepsilon(x)\left(2F_{a}^{~c}F_{bc}-\frac{1}{2}g_{ab}F_{cd}F^{cd}\right), 
\end{equation}
represents the energy-momentum tensor with its scalar sector $\Theta_{ab}$ presented in Eq. \eqref{emt}. The action \eqref{1action} is not invariant under diffeomorphisms, presenting explicit symmetry breaking in both scalar and Maxwell sectors induced by the background dependence. It causes a compatibility problem with the gravitational field equation \eqref{ee3} since we must have $\nabla_a\mathcal{E}^a_{~b}=0$ but each nondynamical degree of freedom within the action \eqref{1action} provides a new contribution to the gradient of the energy-momentum tensor \eqref{emt1}, while the contracted Bianchi identity $\nabla_a G^{ab}=0$ holds even in the absence of the general covariance due to the Riemann tensor symmetries. It restricts the possibility of existence of solutions to cases where the sum of contributions arising from the background dependence comes in such a way that $\nabla_a T^a_{~b}=0$, which leads to 
\begin{equation}\label{boundeq}
    \partial_a V(x, \phi)=-\frac{1}{2}F_{cd}F^{cd}\partial_a \varepsilon(x), 
\end{equation}
used as a compatibility equation. 

The {\it ansatz} we use for the background geometry is given by the metric \eqref{backmetric}, but for simplicity we focus on cases where the metric functions are related by $\nu(r)=-\lambda(r)$. Motivated by the results described in the previous section we assume that the scalar potential is
\begin{equation}\label{1potmodel}
V(r,\phi)=\frac{1}{2}e^{-2\nu}\left(\frac{\ell}{r}\right)^{2(D+z-2)}\left(\frac{dW}{d\phi}\right)^2+U(r),
\end{equation}
for some functions $W=W\left(\phi\right)$ and $U(r)$ to be specified. In this way, we factorize the dynamical degrees of freedom from the nondynamical ones without affecting the field equation \eqref{ee1}, which can still be reduced to the first-order equation given by \eqref{1ordereq}. Furthermore, we require the effective dielectric and both scalar and Maxwell fields to have again only radial dependence, i.e.,
\begin{equation}
\phi=\phi(r), ~~A=A(r)dt~~\text{and}~~\varepsilon(x)=\varepsilon(r).
\end{equation}
In particular, the Maxwell field equation \eqref{ee2} becomes
\begin{equation}
    A'(r)=\frac{q/\ell}{\varepsilon(r)}\left(\frac{\ell}{r}\right)^{D-z-1},
\end{equation}
where $q$ denotes an integration constant and prime denotes derivation in relation to the $r$-coordinate. We can associate to the Maxwell field a charge given by
\begin{eqnarray}
Q=-\frac{1}{4\pi}\!\oint_{\partial\Sigma} \!\!\!d^{\small{D-2}}x \sqrt{|h^{(2)}|} n_a s_b \widetilde{F}^{ab}=\frac{q}{4\pi\ell}\omega_{D-2}^{(\gamma)},~~~~~
\end{eqnarray}
where $\widetilde{F}^{ab}=\varepsilon (x)F^{ab}$. One can also verify that the compatibility equation \eqref{boundeq} is satisfied by the pair of functions
\begin{subequations}
\begin{eqnarray}
    \frac{1}{\varepsilon(r)}&=&\frac{\left(z-1\right)\left(D+z-2\right)}{2q^2}\left(\frac{r}{\ell}\right)^{2(D-2)}e^{2\nu},~~\\[2pt]
    U(r)&=&-\frac{z\left(z-1\right)}{2\ell^2}e^{2\nu},\label{ufunc}
\end{eqnarray}
\end{subequations}
and note that both approach zero in the relativistic limit $z\to 1$. Note that these background geometry dependencies affect obtaining the expression for the Maxwell field, which can now only be obtained after solving the gravitational field equations \eqref{ee3}. The components $(r,r)$ and $(t,t)$ of the zero tensor $\mathcal{E}^a_{~b}$ in Eq. \eqref{ee3} becomes
\begin{subequations}
\begin{eqnarray}
\nonumber\!\!\!\mathcal{E}^{r}_{~r}\!&=&\Lambda-\frac{\left(D-2\right)\left(D-3\right)\gamma}{2\ell^2r^2}+\frac{\left(D-2\right)\left(z-1\right)}{\ell^2}e^{2\nu}\\[2pt]
\label{eer}&~&\!\!-\frac{e^{2\nu}}{2}\phi'^2\!\left(\frac{r}{\ell}\right)^{\!2}\!\!+\!\frac{D-2}{2\ell^2 r^{D+2z-4}}\!\left(r^{D+2z-3}e^{2\nu}\right)'\!=0,~~~~~~~~~\\[4pt]
\nonumber\mathcal{E}^{t}_{~t}&=&\Lambda-\frac{\left(D-2\right)\left(D-3\right)\gamma}{2\ell^2r^2}+\frac{\left(D-2\right)\left(z-1\right)}{\ell^2}e^{2\nu}\\[2pt]
\label{eet}&~&+\frac{1}{2}e^{2\nu}\phi'^2\left(\frac{r}{\ell}\right)^2+\frac{D-2}{2\ell^2 r^{D-2}}\left(r^{D-1}e^{2\nu}\right)'=0.
\end{eqnarray}
\end{subequations}
In the above equations we have no charge term contributions associated with the Maxwell field. This comes from the compatibility Eq. \eqref{boundeq}, which means that the effective dielectric, the scalar and Maxwell fields compete in such a way that the background geometry remains uncharged. The remaining equations $\mathcal{E}^i_{~j}=0$ are left for checking. From subtracting the identity \eqref{eet} from \eqref{eer} one finds
\begin{equation}
    \left(\frac{d\phi}{dr}\right)^2=\frac{\left(D-2\right)\left(z-1\right)}{r^2},
\end{equation}
which is used within Eq. \eqref{eet} to finally find the background solution
\begin{equation}\label{metricsol}
    e^{2\nu}=1-2m\left(\frac{\ell}{r}\right)^{D+3z-4}+\frac{\gamma_z}{r^2},
\end{equation}
with $\gamma_z=\left(D-3\right)\gamma/\left(D+3z-6\right)$ and $m$ is an integration constant interpreted as a ``mass parameter'', along with the field solutions
\begin{subequations}
\begin{eqnarray}
        \phi(r)&=&\varphi\pm\sqrt{\left(D-2\right)\left(z-1\right)}\ln \frac{r}{\ell},\\[3pt]
  \nonumber  A(r)&=&\Phi+\frac{\left(D+z-2\right)m}{2q}\left(\frac{\ell}{r}\right)^{2(z-1)}+\\[3pt]
    &~&+\frac{z-1}{2q}\!\left(\frac{r}{\ell}\right)^{\!D+z-2}\!\left(1+\frac{\left(D+z-3\right)\gamma_z}{\left(D+z-4\right)r^2}\right)\!,~~~~~~~~~
    \end{eqnarray}
\end{subequations}
where the pair $(\varphi,\Phi)$ are integration constants. Since the scalar field solution is invertible, the auxiliary function $W(\phi)$ is found by using Eq. \eqref{1ordereq}, leading to
\begin{eqnarray}
\nonumber W(\phi)&=&W\left(\chi(\phi)\right)\\[2pt]
\nonumber &=&\frac{(D-2)(z-1)}{\ell}\left(\frac{\chi^{D+z-2}}{D+z-2}+\frac{m\chi^{-2(z-1)}}{z-1}\right.\\[2pt]
\label{wphi}&~&\left.+\frac{\gamma_z}{\ell^2}\frac{ \chi^{D+z-4}}{(D+z-4)}\right)
\end{eqnarray}
with $\chi(\phi)=\exp\left(\pm\left(\phi-\varphi\right)/\sqrt{(D-2)(z-1)}\right)$.  Once we know the expression for $W(\phi)$ its simple to find the scalar potential by inserting expressions \eqref{wphi}, \eqref{metricsol} and \eqref{ufunc} in  Eq. \eqref{1potmodel}. For $z=1$ the auxiliary function, the scalar and Maxwell fields becomes constants, while the effective dielectric and the function $U(r)$ goes to zero, which implies that in this regime, in addition to recovering the AdS solution, we cancel all contributions arising from the diffeomorphism invariance breaking, restoring the general covariance. 
\section{Thermodynamics}
\begin{figure*}[t!]
  \centering
  \begin{tabular}{@{\hspace*{0.01\linewidth}}p{0.5\linewidth}@{\hspace*{0.025\linewidth}}p{0.45\linewidth}@{}}
  	\centering
   	\subfigimg[scale=0.38]{(a)}{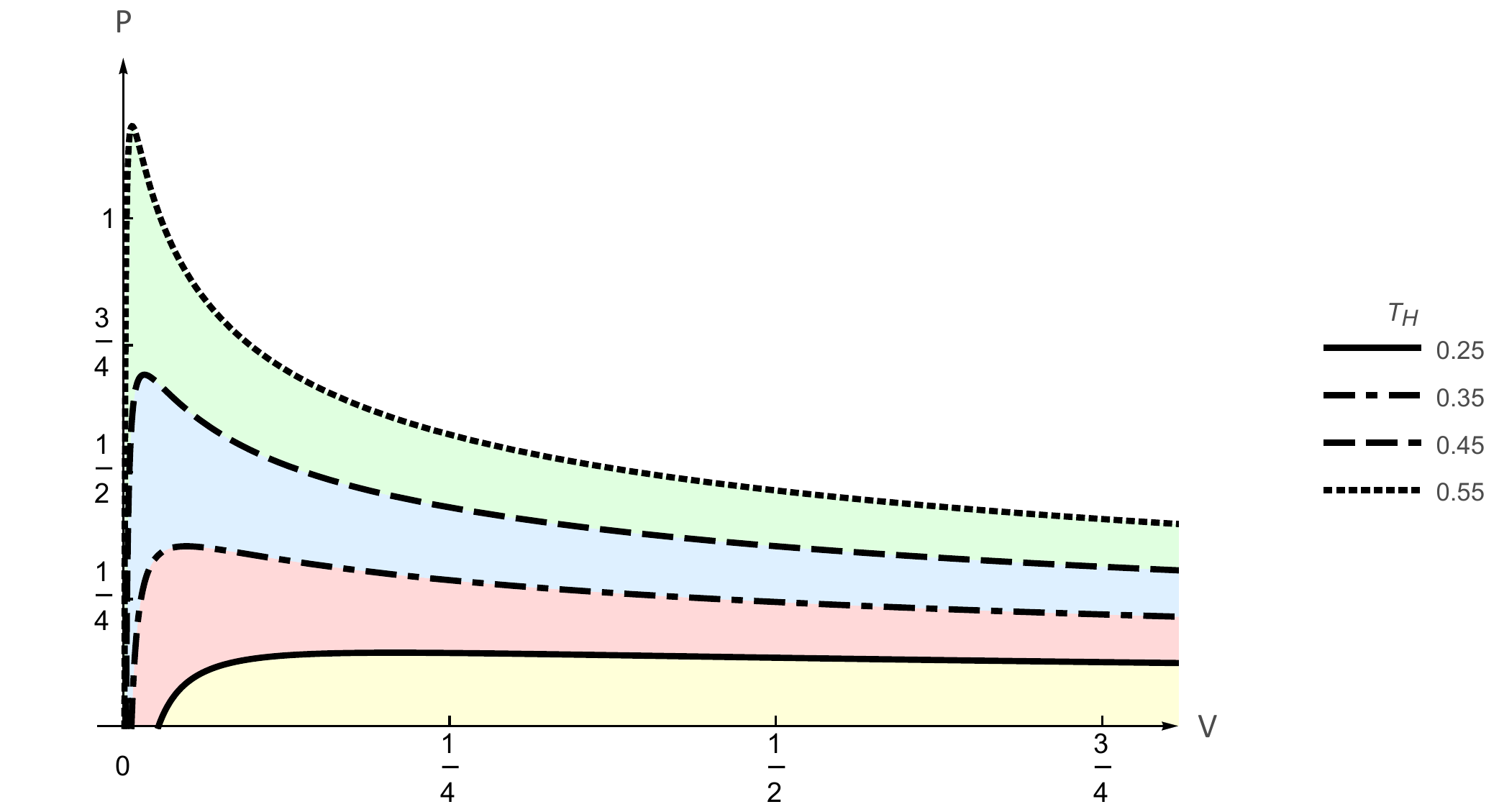} &
   	\subfigimg[scale=0.38]{(b)}{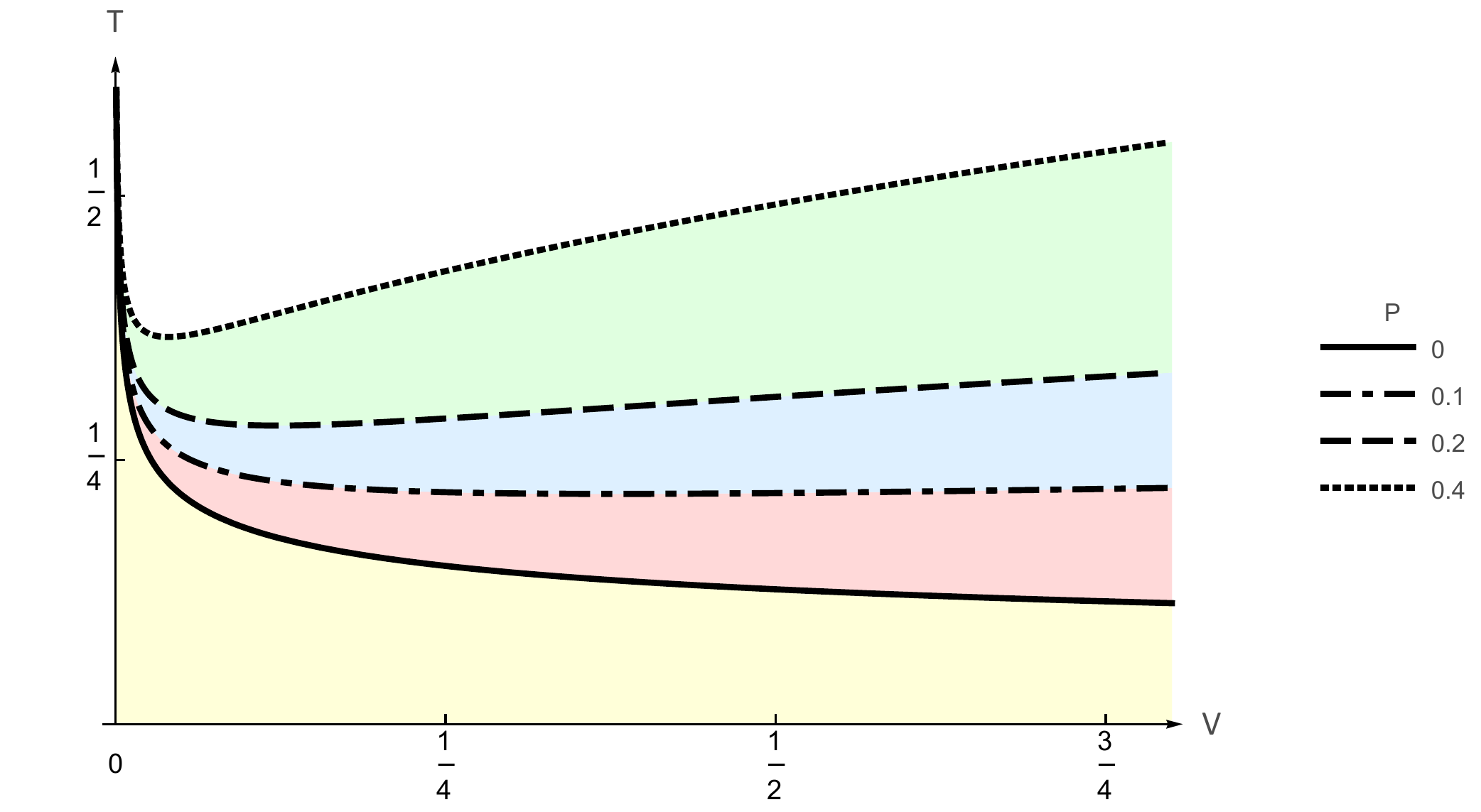}
  \end{tabular}
  \caption{(a) $P-V$ diagram for isotherms with different values of $T_H$ and (b) $T-V$ diagram for curves of  constant pressure from Eq. \eqref{th} with $\ell=1,~\gamma=1,~z=1.25$ and $D=3+1$.}
	\label{fig2}
\end{figure*}
\begin{figure}[t]	
		\centering
\includegraphics[width=8cm]{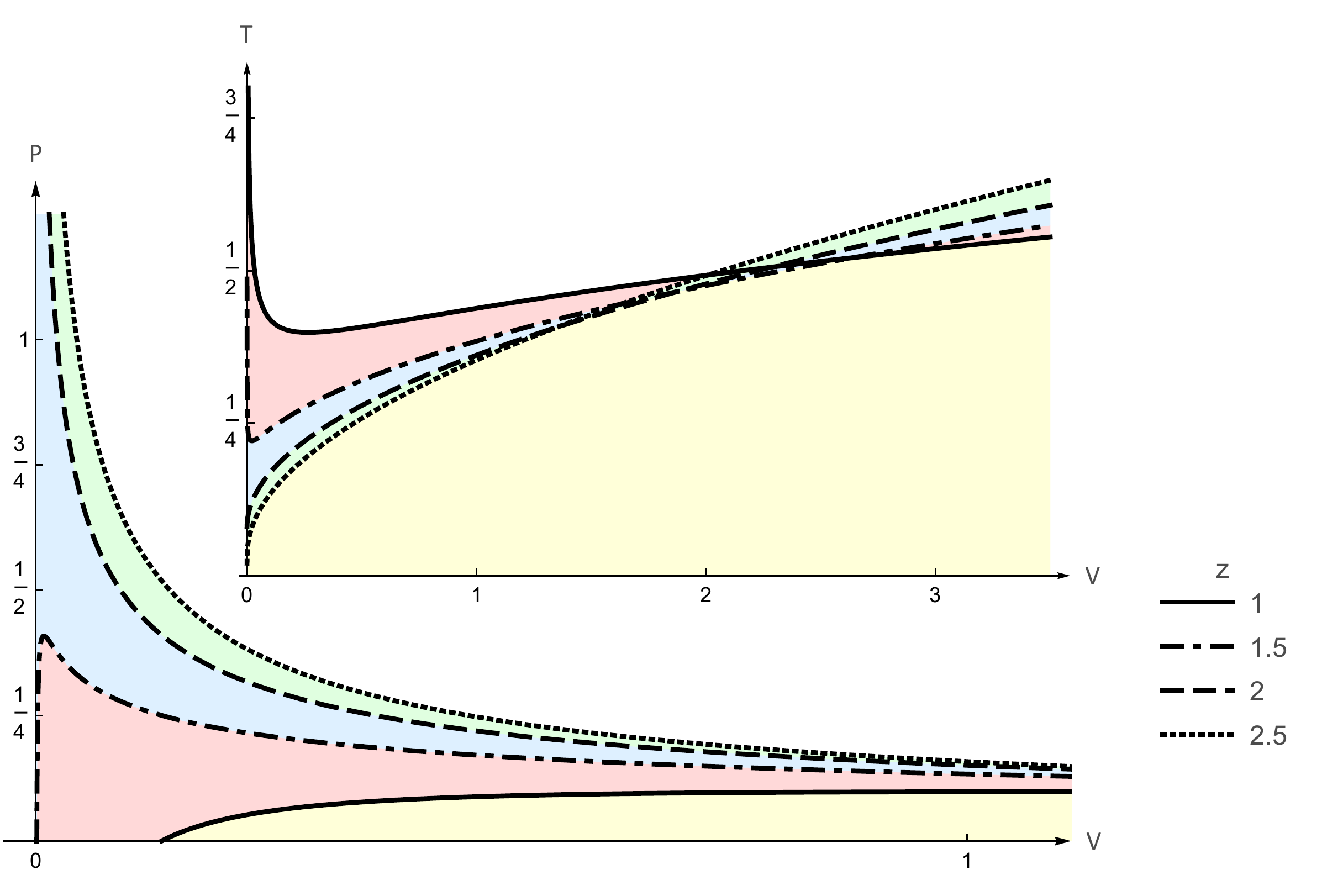}\\
\vspace{2mm}
\caption{$P-V$ (with fixed $T_H=0.25$) and  $T-V$ (with fixed $P=0.25$) diagrams for Eq.\eqref{th} with $\ell=1,~\gamma=1$, $D=3+1$ and some different values of $z$.}
\label{fig3}
\end{figure}

We choose the extended thermodynamics framework \cite{dolan2011cosmological,kubizvnak2017black} for the analysis of black hole thermodynamics obtained in the previous section. The cosmological constant is understood as a constant thermodynamic pressure term given by $P=-\Lambda/8\pi$ 
\cite{kastor2009enthalpy} and the mass of the black hole is read as the enthalpy of the solution, rather than its internal energy. The black hole temperature and entropy are given by
\begin{subequations}
\begin{eqnarray}
    T_{H}&=&\frac{1}{4\pi \ell}\left(\frac{r_h}{\ell}\right)^{\!z}\left(\left(D+3z-4\right)+(D-3)\frac{\gamma}{r_h^2}\right),~~~~~~~\\[3pt]
    S_{bh}&=&\frac{1}{4}\left(\frac{r_h}{\ell}\right)^{D-2} \omega^{\left(\gamma\right)}_{D-2},
\end{eqnarray}
\end{subequations}
respectively, while the first law becomes
\begin{eqnarray}
\label{firstlaw}dM&=&T_{H}dS_{bh}+V_{bh}dP,
\end{eqnarray}
where $V_{bh}$ denotes the thermodynamic volume conjugate to the pressure. 

The black hole thermodynamic mass is obtained by integrating equation \eqref{firstlaw} at constant pressure from zero to the horizon and can be expressed in terms of pressure and entropy  as 
\begin{eqnarray}
M(P,S_{bh})=\frac{S_{bh}}{4\pi\ell}\left(\frac{\hat{\gamma}\widetilde{S}_{bh}^{\frac{z-2}{D-2}}}{D+z-4}+\frac{16\pi\ell^2 P\widetilde{S}_{bh}^{\frac{z}{D-2}}}{D+z-2}\right),~~~~~
\end{eqnarray}
where, for simplicity, we define $\widetilde{S}_{bh}=4S_{bh}/\omega_{D-2}^{(\gamma)}$. A direct check reveals that $T_{H}=\left.\left(\partial M/\partial S_{bh}\right)\right|_P$ and since the black hole mass is identified with enthalpy, the thermodynamic volume becomes
\begin{eqnarray}
V_{bh}=\left.\frac{\partial M}{\partial P}\right|_{S_{bh}}\!\!=\frac{\ell}{D+z-2}\left(\frac{r_h}{\ell}\right)^{D+z-2}\omega_{D-2}^{(\gamma)}, ~~~
\end{eqnarray}
which naturally matches the Smarr-like relation 
\begin{eqnarray}
\label{smarr}M&=&\frac{D-2}{D+z-4}T_H S_{bh}-\frac{2}{D+z-4}PV_{bh},
\end{eqnarray}
also found in \cite{dayyani2018critical} for another class of Lifshitz black holes, which coincides with the known Smarr relation associated with AdS black holes for $z=1$ \cite{dolan2011cosmological,brenna2015mass}. In this way, the internal energy is Legendre transform of the mass,
\begin{eqnarray}\label{eint}
    E_{int}=M\!-\!PV_{bh}=\frac{\hat{\gamma}\omega_{D-2}^{(\gamma)}}{16\pi\ell\left(D+z-4\right)}\left(\frac{r_h}{\ell}\right)^{D+z-4},~~~~~
\end{eqnarray}
which is proportional to the curvature on the horizon.  

By assuming that the temperature can be expressed as a function of the pressure and volume, the state equation for the system is
\begin{eqnarray}\label{th}
T(P,V_{bh})=\frac{\widetilde{V}_{bh}^{\frac{z}{D+z-2}}}{4\pi\ell(D-2)}\left(16\pi\ell^2 P+\hat{\gamma}\widetilde{V}_{bh}^{\frac{-2}{D+z-2}}\right),~~
\end{eqnarray}
with $\widetilde{V}_{bh}=(D+z-2)V_{bh}/\ell\omega_{D-2}^{(\gamma)}$. Its associated $P-V$ an $T-V$ diagrams for fixed $z$ are depicted in Fig. (\ref{fig2}-a) and Fig. (\ref{fig2}-b), respectively. In Fig.\eqref{fig3} we depicted the $P-V$ and $T-V$ diagrams with fixed $T$ and $P$, respectively, for different values of $z$. The specific heat at constant pressure is $C_p=T_H/\left.\left(\partial T_H/\partial S_{bh}\right)\right|_P$, which leads to
\begin{eqnarray}
    C_p=(D-2)S_{bh}\frac{16\pi\ell^2P\widetilde{S}_{bh}^{\frac{2}{D-2}}+\hat{\gamma}}{16\pi\ell^2 z P\widetilde{S}_{bh}^{\frac{2}{D-2}}+\hat{\gamma}(z-2)}. 
\end{eqnarray}
For $\hat{\gamma}=0$ the internal energy \eqref{eint} is zero and the temperature is simply $T=4\ell P\widetilde{V}^{\frac{z}{D+z-2}}/(D-2)$, while the specific heat is proportional to the entropy, with $C_p=(D-2)S_{bh}/z$, which in this case is always positive (Note that for $(\hat{\gamma},z,D)=(0,2,4)$ the entropy and specific heat are equal). In addition, for $\hat{\gamma}\neq0$ we have $C_p>0$ only for $-2\Lambda r_h^2>|\hat{\gamma}|$, which sets a lower bound on the event horizon possible values,
\begin{equation}
    r_h>\sqrt{\frac{D-3}{D+3z-4}}\ell.
\end{equation}
These results generalize those obtained in \cite{dolan2011cosmological} in the Ads setup for the Lifshitz black holes we present here. After some manipulations, the thermodynamic mass can be also expressed in terms of the ``mass parameter'' $m$ as
\begin{eqnarray}
    M=\widetilde{M}-\frac{(z-1)(D-2)\omega_{D-2}^{(\gamma)}}{4\pi\ell(D+z-2)(D+z-4)}\!\left(\frac{r_h}{\ell}\right)^{\!D+z-2},~~~~~~
\end{eqnarray}
with 
\begin{eqnarray}
\widetilde{M}=\frac{D+3z-6}{D+z-4}\!\left(\frac{\ell}{r_h}\right)^{\!2(z-1)}\frac{(D-2)\omega_{D-2}^{(\gamma)}}{8\pi\ell}m,~~~~~~
\end{eqnarray}
proportional to the mass of a topological AdS$_{(D)}$ black hole. For $z=1$ we have $M=\widetilde{M}_{(z=1)}$, which coincides with the mass of the topological AdS$_{(D)}$ \cite{chamblin1999charged,banerjee2011thermodynamics}, as expected. 
\section{Ending comments}
In this work we studied scalar fields in asymptotically Lifshitz spacetimes in scenarios presenting general covariance breaking from the requirement of coordinate dependence within the action. In the probe limit we presented a method to capture analytical and numerical spatially localized solutions based on evasions of the Derrick's theorem in curved backgrounds, generalizing the previous results of \cite{moreira2022analytical}. The energy-momentum tensor is not conserved in these setups, but when considering both background geometry and field solutions static, we keep the main conditions to have finite energy, as expected from Noether's theorem. The solutions found are radially stable and a zero mode can be captured from its stability equation. 

We also extended our study to consider backreaction effects from an Einstein-scalar-Maxwell model presenting explicit coordinate dependence in both scalar and Maxwell sectors. In the scalar sector - inspired by the expression we found for the scalar potential in the probe limit - we factorized the scalar sector into two subsectors, one depending only on the scalar field and the other holding nondynamical degrees of freedom, while the Maxwell sector was equipped with an effective dielectric, also coordinate dependent. In this setup the diffeomorphism invariance breaking induces a compatibility problem with Einstein equations, since in general the energy-momentum tensor is no longer conserved while the Einstein tensor still holds zero divergence due to the Bianchi identity. We can only solve the complete set of field and gravity equations in the particular case where the sum of the contributions arising from the nondynamical degrees of freedom arising in the divergence of the energy-momentum tensor is zero, which was used to specify the background geometry of the model, revealing a noncharged Lifshitz black hole solution whose thermodynamics we analyzed in detail.

An interesting feature of the solution we found is that the effects arising from the general covariance breaking elements inserted in the action disappear in the relativistic limit $z\to 1$, despite the fact that there was no required condition relating these two ingredients in the initial assumptions. Furthermore, the emerging compatibility equation exhibits a competition between the effective dielectric and the scalar potential which has the net effect of discharging the background geometry. We must point out that there are several paths to extend the ideas outlined here. For example, new solutions can be found when considering distinct effective actions or different geometries presenting anisotropic scaling, such as those discussed in \cite{pedraza2019hyperscaling}. Another possibility is the search for new gravity solutions in geometries with no scaling anisotropy - the probe limit in these cases has already been explored in \cite{moreira2022localized}, but scenarios which present backreaction have not yet been considered. Finally, we hope that the idea of using minimal energy solutions on probe limits in a coordinated way with explicit general covariance breaking will be useful for other future works on effective gravitational models and their applications, since the variety of choices which can be made about diffeomorphism invariance breaking brings us a wide range of possibilities to explore.

\begin{acknowledgments}
D.C.M. would like to thank the Brazilian agencies CNPq and FAPESQ-PB for the financial support (PDCTR FAPESQ-PB/CNPq, Grant no. 317985/2021-3). F.A.B. acknowledges support from CNPq (Grant nos. 312104/2018-9, 439027/2018-7) and also CNPq/PRONEX/FAPESQ-PB (Grant nos. 165/2018), for partial financial support.
\end{acknowledgments}
\bibliography{biblio} 
\end{document}